**Synthesis and In Situ Modification of Hierarchical SAPO-34 by PEG with Different Molecular Weights; Application in MTO Process**

Seyed Hesam Mousavi[a], Shohreh Fatemi[a,*], Marjan Razavian[a]

[a] *School of Chemical Engineering, College of Engineering, University of Tehran, Tehran, Iran, P.O. Box: 11365 4563*

**Abstract**. Modified structures of SAPO-34 were prepared using polyethylene glycol as the mesopores generating agent. The synthesized catalysts were applied in methanol-to-olefins (MTO) process. All modified synthesized catalysts were characterized via XRD, XRF, FESEM, FTIR, $N_2$ adsorption-desorption techniques, and temperature-programmed $NH_3$ desorption and they were compared with conventional microporous SAPO-34. Introduction of non-ionic PEG capping agent affected the degree of homogeneity and integrity of the synthesis media and thus reduced the number of nuclei and order of coordination structures resulting in larger and less crystalline particles compared with the conventional sample. During the calcination process, decomposition of absorbed PEG moieties among the piled up SAPO patches formed a great portion of tuned mesopores into the microporous matrix. These tailored mesopores were served as auxiliary diffusion pathways in MTO reaction. The effects of molecular weight of PEG and PEG/Al molar ratio on the properties of the synthesized materials were investigated in order to optimize their MTO reaction performance. It was revealed that both of these two parameters can significantly change the structural composition and physicochemical properties of resultant products. Using PEG with MW of 6000 has led to the formation of RHO and CHA structural frameworks i.e. DNL-6 and SAPO-34, simultaneously, while addition of PEG with MW of 4000 resulted the formation of pure SAPO-34 phase. Altering the PEG/Al molar ratio in the precursor significantly influenced the porosity and acidity of the synthesized silicoaluminophosphate products. SAPO-34 impregnated with PEG molecular weight of 4000 and PEG/Al molar ratio of 0.0125 showed superior catalytic stability in MTO reaction because of the tuned bi-modal porosity and tailored acidity pattern. Reactivation experiments also revealed that mesopores were stably preserved even after several regeneration cycles.

* Corresponding author: Tel: +98 21 61112229; Fax: +98 21 669667784.
  E-mail address: shfatemi@ut.ac.ir (Sh. Fatemi).

## 1. Introduction

Zeolites are the most widely used solid catalysts in industry due to their microporous structures and the resulting unique shape selectivity [1]. Silicoaluminophosphate-34 (SAPO-34) molecular sieve with the topological structure of chabazite (CHA) has been intensively studied as one of the most excellent catalysts for the methanol-to-olefin (MTO) reaction. Such process provides an alternative route to produce light olefins from nonpetroleum sources, and it has gained considerable attention in the last three decades [2].

Three-dimensional pore system of SAPO-34 molecular sieve consists of large cavities (6.7×10 Å) separated by small windows (3.8×3.8 Å). Periodic building unit is the double 6-ring layer connected through 4-rings. In comparison with ZSM-5, the small entrances of SAPO-34 molecular sieve cages located within 8-membered-rings (8-MR) only allow the diffusion of linear hydrocarbons [3], [4]. Since aromatics and branched hydrocarbons are too large to transfer through its 8-MR pore entrances, the light olefins are naturally predominant in product streams. In addition, the cage shape and size of SAPO-34 molecular sieve has been found to be suitable to preserve reactive intermediates that produce lower olefins selectively [5]. Therefore, in the MTO reaction over the SAPO-34 catalyst, high methanol conversion and high selectivity to light olefins can be achieved due to relatively moderate acidic strength, and shape selectivity of porous structure [3]. However, the SAPO-34 catalyst shows rapid deactivation due to coke formation and active sites encapsulation, which completely blocks the internal channels of the structure during MTO reaction. This phenomenon inhibits the access of reagents to active sites located within pores and wear out a huge part of internal structure. Consequently the catalytic performance deteriorates and catalyst deactivates within a short period of time. This coke formation is mainly caused by the diffusion restriction of reactants and products in small zeolite pores, and it has severely limited application of SAPO-34 in MTO process [6], [7].

Diverse attempts ranging from the synthesis of nano-sized zeolites [8], zeolitic-ordered mesoporous materials [9], zeolites with a secondary porosity created via steaming [10], desilication [11] or based on the confined space synthesis protocol [12] and delaminated zeolites [13] have been reported as solutions to the quick deactivation problem. Regarding these efforts, hierarchical zeolitic structures are designated as zeolites with a bi/tri-modal pore size distribution, which consequently exhibit reduced steric and diffusional restrictions [14].

Manipulating the nanocrystals into hierarchical nanoarchitectures using a self-assembly strategy is an efficient way to obtain structured catalysts with higher activities, longer lifetimes and better selectivities [15]–[18].

In recent decades, some methods have been successfully employed for the preparation of hierarchical porous aluminosilicate zeolites including post-treatment [19], hard templating [20], and soft-templating methods [21]. Notably, Ryoo's group demonstrated the soft-templating synthesis of zeolites by introducing organosilane surfactant as the mesoporosity directing agent into conventional zeolite synthesis precursor [22]. They further extended this strategy to the synthesis of mesoporous aluminophosphate molecular sieves [23].

However, the development of an effective strategy for the formation of hierarchical SAPO-34 with excellent performance in MTO reactions is still a great challenge for the materials, chemical engineering, catalysis, and particuology communities.

In this contribution, SAPO-34 molecular sieve with tuned hierarchical nanostructure was synthesized through a facile one-steps hydrothermal route using a combination of amine agents [i.e., tetraethyl ammonium hydroxide (TEAOH), diethyl amine (DEA)] and polyethylene glycol (PEG), as structure directing and mesogenerating agents, respectively. Our aim was focused on the introduction of tuned transport pores in the meso scale to advance the molecular diffusion property and enhance the accessibility of active sites within the cages and reduce the intervention of small micropores in the diffusion of reactants and products. This method is inexpensive and easily reproduced for high-yield production of SAPO- 34 catalysts with good crystallinity. We further investigated the catalytic performance of the synthesized hierarchical SAPO-34 samples in MTO process, aiming to reveal the intrinsic correlation between the catalytic performance and the accessibility of active sites, as well as the correlation between the accessibility and textural properties. Overall, we discuss here the performances of different types of SAPO-34 with hierarchical structures and compare them with the performance of conventional sample (PEG free synthesis) to attain a broad picture for better understanding of the role of tailored structure in MTO process.

## 2. Experimental

### 2.1. Synthesis of catalysts

SAPO-34 molecular sieves were synthesized from a reaction mixture with molar composition of 1 $Al_2O_3$: 0.5 $SiO_2$: 0.8 $P_2O_5$: 0.1 HCl: 1.8 DEA: 0.2 TEAOH: (x, MW) PEG: 60 H2O wherein x indicates the PEG/Al molar ratio and MW shows the molecular weight of the used PEG. First, aluminum isopropoxide (98%, Merck) was dissolved in mixed acid solution (HCl: 37%, aqueous solution, Merck and phosphoric acid: 85%, aqueous solution, Merck) step by step under vigorous stirring. Then tetraethyl orthosilicate (98%, Merck) was dissolved in basic templates (TEAOH: 20%, Merck and DEA: 99%, Merck) solution smoothly and was further added to aluminum solution. As the next step, polyethylene glycol (MW=4000 and 6000, Merck) was added and the resultant gel was allowed to age and hydrolyze at room temperature for 24 h. Finally, the precursor gel was hydrothermally treated at 130℃ and 200℃ for 5 and 12 h, respectively. The solid product was recovered by centrifugation and washed four times, and then dried at 100 ºC. Calcination was performed at 550 ºC for 5 h to remove the organic templates molecules from materials backbones.

For comparison, conventional SAPO-34 particles were synthesized through the same procedure without using PEG soft template. Table 1 shows the molecular weight of used PEG and molar ratio of PEG to $Al_2O_3$ in the synthesized samples.

**Table 1**

Synthesis condition for the as-synthesizes hierarchical silicoaluminophosphate catalysts.

| Sample name | PEG MW | PEG/Al molar ratio |
|---|---|---|
| 4-0.025 | 4000 | 0.025 |
| 4-0.0125 | 4000 | 0.0125 |
| 6-0.025 | 6000 | 0.025 |
| 6-0.0125 | 6000 | 0.0125 |
| conventional | - | - |

### 2.2. Characterization

X-ray powder diffraction (XRD) patterns were obtained by D8 ADVANCE X-Ray (Bruker AXS) diffractometer (Cu Kα radiation) in the 2tetha range of 2–50. Chemical composition of the solid samples was determined by using a SPECTRO XEPOS X-ray fluorescence (XRF) spectrometer. Field emission scanning electron microscope (FESEM) images were recorded with Hitachi-S4160 microscope operating at 30 kV. $N_2$ adsorption-desorption isotherms were recorded using Bel-SORP MINI II apparatus. Textural properties including total and micropore surface areas and volumes were measured by BET (Brunauer–Emmett–Teller) and t-plot methods. Micropore and mesopore size distributions were analyzed by MP-plot (micropore analysis) and BJH (Barrett– Joyner–Halenda) methods. Before the surface area measurements, samples were degassed at 300 °C for 3 h. Fourier transform infrared (FTIR) spectra were carried out using Perkin-Elmer spectrometer in the frequency range of 4000-400 $cm^{-1}$. Temperature-programmed desorption (TPD) measurements were carried out on a conventional apparatus using BELCAT-B analyzer. In a typical TPD measurement, a 0.10 g sample was pretreated at 500 ℃ for 1 h under flowing Helium to remove adsorbed water. The sample was then saturated with $NH_3$ at 100 ℃ for 1.0 h and purged with He flow at 100 ℃ for 30 min to remove weakly adsorbed $NH_3$. The temperature was then increased at a heating rate 10 ℃/min from 100 to 900 ℃.

### 2.3. MTO reaction tests

MTO reaction was carried out in a fixed-bed reactor at atmospheric pressure. 0.5 g of catalyst was loaded into the stainless steel reactor. The reactor was 1.2 cm in diameter and 50 cm long. The temperature was measured by three thermocouples at above, bottom, and the center of the catalyst bed, and the temperature was controlled by three PID controllers (JOMO) with ±1 ºC. Before the reaction test, the catalyst was pretreated in a flow of $N_2$ at the reaction temperature of 400 ºC for 1 h to activate the molecular sieve. Then a liquid mixture of methanol in water (30 wt %) was fed into the catalyst bed using a syringe pump accompanying with $N_2$ stream (10 ml/min) as a carrier gas. Researchers [24], [25] have shown that olefin selectivity can be increased when the methanol partial pressure is decreased by adding diluents to the feed. From the diluents that could be used for such a purpose, water is the best alternative. Water is more readily available, and its vapor has a relative high heat capacity that is good for carrying away the heat of reaction. The weight hourly space velocity (WHSV) was 3 $h^{-1}$. The reaction products were analyzed with an online GC (gas chromatography YL6100) equipped with HID detector. Methanol conversion and product selectivity are defined by Eqs. (1) and (2), respectively:

$$MeOH\ conversion \qquad (1)$$

$$= \frac{mole\ of\ fed\ MeOH\ -\ mole\ of\ unreacted\ MeOH\ -\ mole\ of\ DME}{mole\ of\ fed\ MeOH} \times 100$$

$$Selectivity\ of\ component\ i = \frac{mole\ of\ component\ i}{mole\ of\ product} \times 100 \qquad (2)$$

## 3. Results and discussion

The XRD patterns of the samples are shown in Fig. 1. The conventional, 4-0.0125, and 4-0.025 samples exhibited identical characteristic peaks corresponding to the SAPO-34 CHA structure [26]; a peak at 2θ~9.5° corresponding to the (0 0 1) crystal face reflection with additional peaks at 13°, 16.1°, 20.7°, and 30.7°. However, the relative intensities of the diffraction peaks of (0 0 1) crystal face (2θ ~9.5°) in the 4-0.0125 and 4-0.025 samples were lower than those of the conventional SAPO-34. The slightly diminished peaks in both 4-0.0125 and 4-0.025 samples reflect lower crystallinity of these samples compared with the conventional one. However, using PEG with molecular weight of 6000 (6-0.0125 and 6-0.025 samples) has led to the formation of RHO framework along with CHA structure. Aluminosilicate zeolite RHO, firstly reported by Robson, is often synthesized in the presence of sodium and cesium cations [27]. The body-centered cubic symmetric structure of RHO is composed of α-cages linking through double 8-rings (D8R) with pore size of 3.6 × 3.6 Å. This structure would undergo a framework distortion with change of unit cell symmetry or even collapse upon extra framework cation exchange or dehydration [28]–[30]. However, the use of expensive cesium cations in the synthesis of RHO zeolite is a limitation on its application. Thus, these interesting results introduce a new and efficient way for synthesizing silicoaluminophosphate molecular sieve with the RHO framework.

The extra peaks at 2θ~7° and 15° [(1 0 0) and (2 1 1) crystal faces] are assumed to be related to the presence of rare DNL-6 (Dalian National Laboratory Number 6) framework which is an isomorphous crystal of the RHO zeolite with a SAPO-based composition. This material was accidentally discovered at 2011 as the intermediate of the phase-transformation process of SAPO molecular sieves employing SAPO-5 as the precursor and DEA as the template [31]. DNL-6, an 8-ring SAPO molecular sieve with an RHO structure possess large α cages and relatively high acid concentration and thermal strength [32]. The framework of DNL-6 compose of high concentration of Si(4Al) environment that can be ascribed to the easy inclusion of the DEA template with small size into large LTA cage in the RHO structure [33]. This structure normally exhibits large surface area, pore volume and high thermal stability.

Several approaches like surfactant-assisted hydrothermal synthesis [34], solvothermal synthesis [31] and dry-gel conversion [33] have been employed to obtain DNL-6 molecular sieve. But each method has to be performed under a narrow range of the starting gel composition and synthesis conditions.

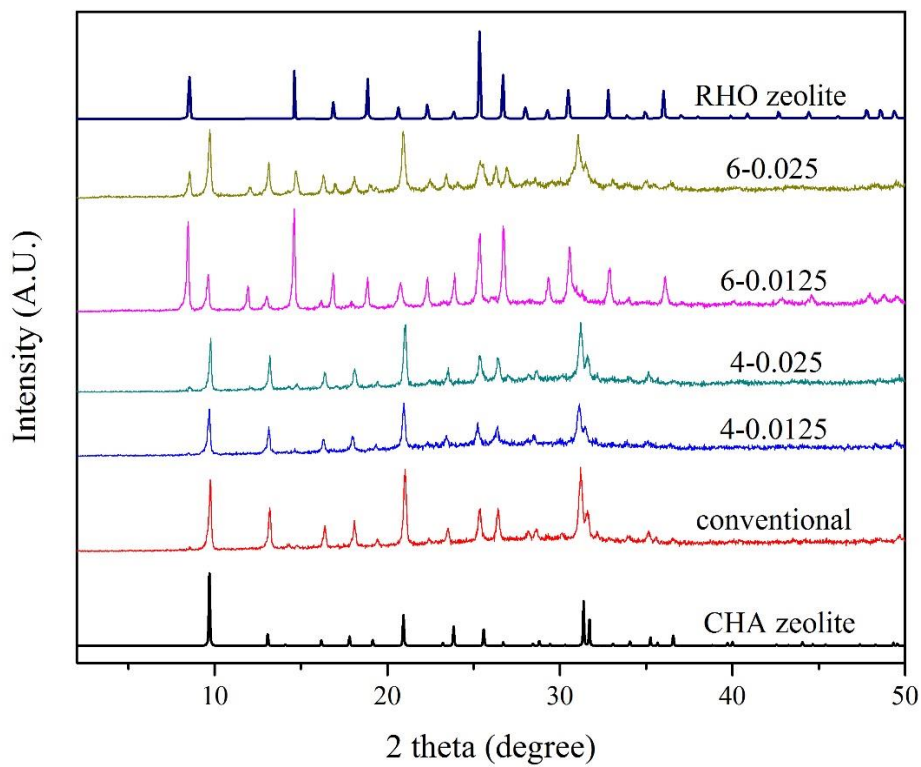

**Fig. 1.** The XRD patterns of the synthesized silicoaluminophosphate catalysts as well as pristine CHA and RHO zeolites [26] as references.

Fig. 2 shows the relative crystallinity of the samples toward CHA and RHO structures. The relative crystallinity was estimated by normalizing against the maximum value attained by summing up the characteristic peak intensities at 2θ = 9.5°, 13°, and 20.7° for CHA framework and 2θ = 6.7°, 15.4°, and 25.3° for RHO framework (Eq. 3).

$$Relative\ Crystallinity\ (\%) = \frac{(I_1 + I_2 + I_3)_{sample}}{(I_1 + I_2 + I_3)_{reference}} \times 100 \qquad (3)$$

Conventional SAPO-34 sample was considered as the reference.

As it can be seen from Fig. 2 and Table 2, by increasing the PEG/Al molar ratio the CHA phase crystallinity is enhanced totally and the formation of DNL-6 with RHO framework is restricted in samples synthesized with 6000MW PEG. Long-chain PEG (MW=6000) acted as a non-ionic capping agent and stabilized DNL-6 embryos providing an appropriate situation for the growth of DNL-6 nucleus. However, DNL-6 is a metasble phase formed within the synthesis system and increasing the acidic PEG amount in 6000-0.025 sample reduced the PH wherein DNL-6 crystals started dissolving in the mother liquor. The dissolution changed the synthesis environment preferring the SAPO-34 crystallization path [34].

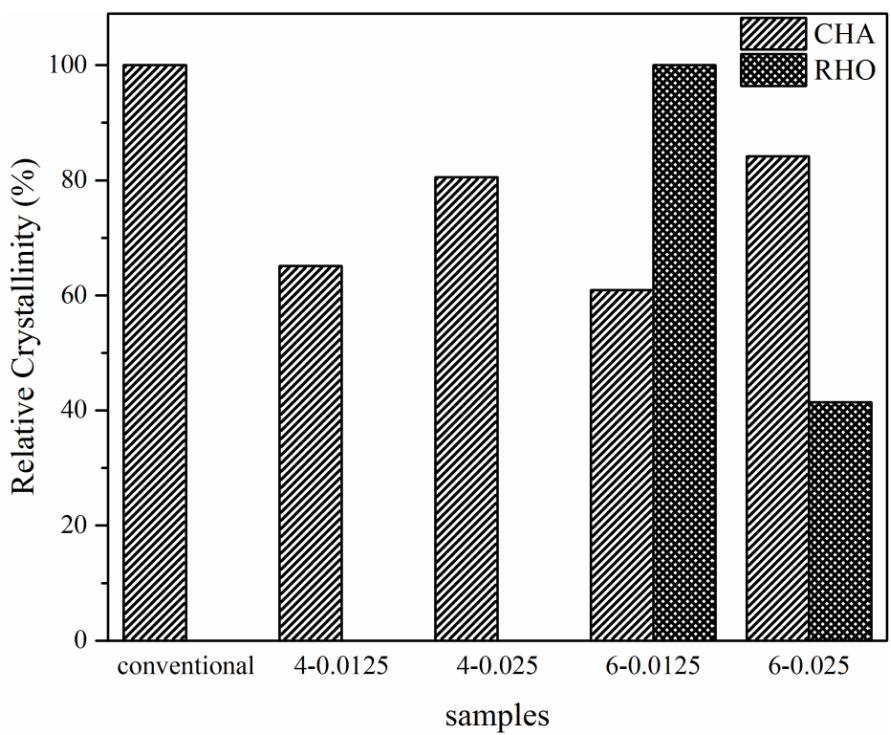

**Fig. 2.** The relative crystallinity of the synthesized silicoaluminophosphate samples toward CHA and RHO frameworks.

**Table 2**

Structural composition of resultant products.

| Sample name | % CHA[a] | % RHO[a] |
|---|---|---|
| conventional | 100 | 0 |
| 4-0.0125 | 100 | 0 |
| 4-0.025 | 100 | 0 |
| 6-0.0125 | 21.5 | 78.5 |
| 6-0.025 | 47.7 | 52.3 |

a) $\% i = \dfrac{(I_1 + I_2 + I_3)_i}{(I_1 + I_2 + I_3)_{CHA} + (I_1 + I_2 + I_3)_{RHO}} \times 100$

Fig. 3 represents the FESEM micrographs of the samples. As shown, conventional sample contained typical semi-cubic shape SAPO-34 crystals. In 4-0.0125 and 0.025 samples, truncated large aggregates composed of rectangular plate like nanocrystals with rugged surfaces are recognized together with some cubic-like particles with more smooth surfaces. Notably, using PEG with molecular weight of 6000 (6-0.0125 and 6-0.025 samples) resulted in the production of two SAPO-34 and DNL-6 structures with different morphologies. Semi-cubic and rhombic dodecahedron particles can be observed in these samples. As evident from Fig. 3 (g-j), a minor fraction with sphere-like features having rough surfaces also coexists which are ascribed to high Si content nanocrystals aggregated spherically [34].

The dissolution of PEG organic additive with high molecular weight reduces the interfacial energy and increase the viscosity and inhomogeneity of the system. According to Van Hook theory the mixing free energy reduces with an increase in viscosity and decrease in surface tension. Therefore, the mixing entropy reduces and phase separation occurs [35], [36]. Consequently, it sounds rational to justify the dominant DNL-6 phase formation in PEG-templated samples (MW=6000) from rich silicon liquid phase through the solution-mediated transport mechanism. In this mechanism the crystals are directly formed from the crystallization of liquid phase, to which the amorphous solid-phase provides nutrition elements through dissolution.

Organic PEG additive increases the viscosity and inhomogeneity of synthesis media, thus inhibiting the transfer of Aluminum and Silicon species. Moreover, the hydroxyl anionic functional groups of ethylene glycol oligomers interact with silanol groups through hydrogen bonding and Al cations by chelating and restrict the mobility of cation moieties. As a result of this low mobility and mass transfer, effective supply of aluminum and silicon is prohibited and nucleation is suppressed. Therefore, lower numbers of nucleus and subsequently larger particles are obtained in PEG-directed samples compared to the conventional one. Furthermore, the aforementioned interactions between end sides of PEG compound and metal cations decrease the symmetry and order of cations coordination structure and result in lower crystallinity of samples which is consistent with XRD results [35], [37], [38].

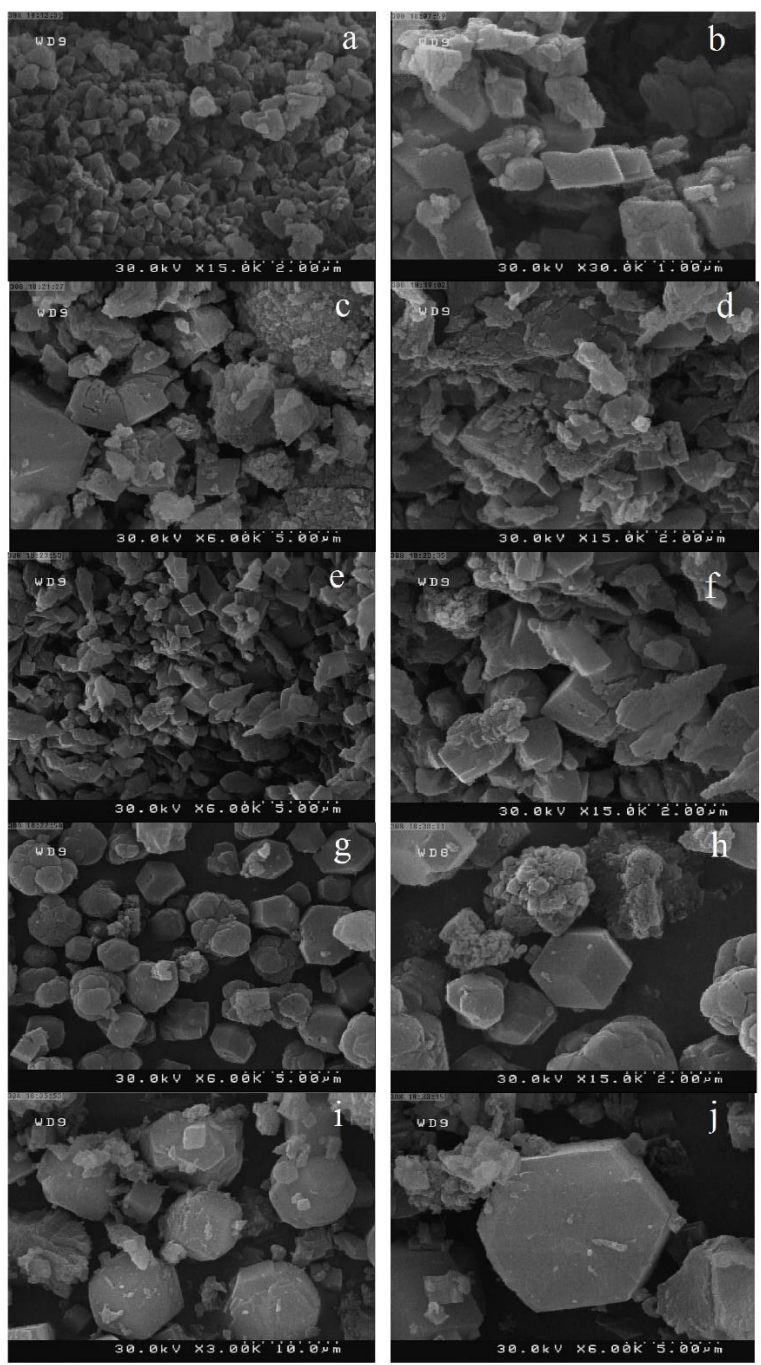

**Fig. 3.** FESEM images of the synthesized silicoaluminophosphate zeolites: the low and high magnification FESEM image of (a, b) conventional, (c, d) 4-0.0125, (e, f) 4-0.025, (g, h) 6-0.0125, and (i, j) 6-0.025.

Another phenomenon which can explain the larger size of resultant particles in PEG-directed samples is the attraction and adsorption of initially formed zeolite units on the surface of PEG micelles. There is a general feeling that PEG is essentially an inert hydrophilic polymer. But indeed, it appears to possess at least some hydrophobic character since at the air-water interface it can form a monolayer and reduce the surface tension, a property which is normally observed using an amphiphilic surfactant that have distinct hydrophilic and hydrophobic parts [39]. When PEG is dissolved in water, SAPO patches will adsorb on small PEG micellar clusters. This occurs because of increasing attraction between the PEO chains. In an earlier direct measurement of the forces between chained PEO groups using the SFA technique, Claesson et al. [40] has found an attractive (adhesion) force between them which increased with temperature rising; but their results suggested that this was mostly the consequence of the reduction in the magnitude and range of the repulsive forces between the PEO groups rather than to an enhancement of the attractive forces. In conventional sample, SAPO units with partial negative charge receiving from substitution of phosphorus atoms of neutral $AlPO_4$ framework with silicon ones repel each other due to electrostatic repulsion resulting in non-aggregated small crystals. In contrast, piled up zeolites on the surface of PEG micelles collide, aggregate and assemble to large self-organized particles especially when the temperature is raised. This way, a great amount of organic chains are embedded within the matrix that leaves cavities upon decomposition of PEG at high temperatures during calcination step. These cavities are served as auxiliary meso transport pores implanted in the backbone of materials.

Fig. 4 demonstrates the FT-IR spectrum of the samples in a frequency range of 400-4000 cm$^{-1}$. There are a broad band at 1098 cm$^{-1}$ assigned to the asymmetric stretching vibration of O-P-O groups. The absorption bands at 490, 640 and 730 cm$^{-1}$ are typical characteristic bands of silicoaluminophosphate framework and attributed to the T-O bending of SiO$_4$ groups and double six-membered ring (D6R) subunits and symmetric stretching of P-O (Al-O) groups. A pronounced shoulder appears around 3435 cm$^{-1}$ in the spectrum which is ascribed to the stretching vibration of hydroxyl groups present in three different structural bridges. These three probable groups consist of Si-OH-Al bridges (Bronsted acid sites), Si/P−OH defects as the result of hydrolysis of Si−O−Si and P−O−Al linkages and terminal Si/P−OH linkages of the external surface of zeolite crystals. This band is notably abated in 4-0.0125 sample which is denotative of the influence of PEG addition on acidic properties.

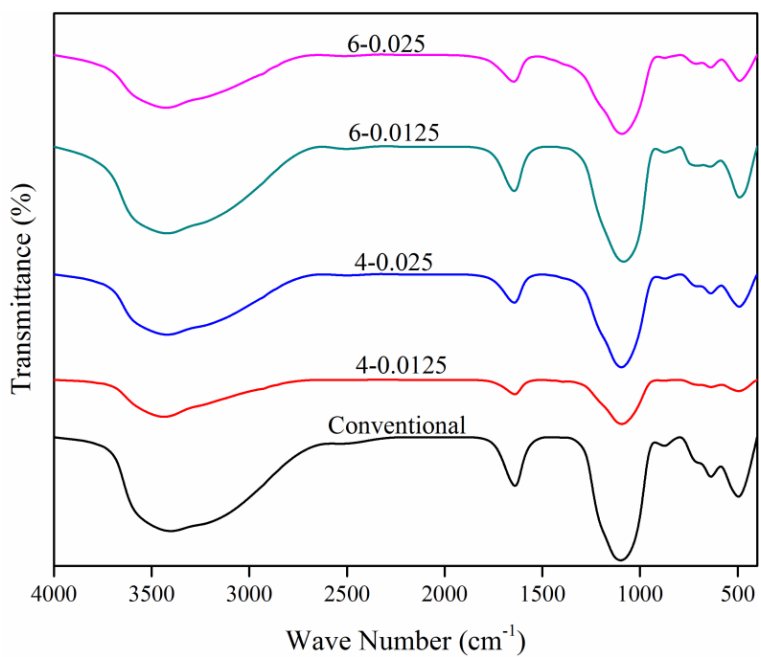

**Fig. 4.** FT-IR patterns of the synthesized silicoaluminophosphate zeolites.

The $N_2$ adsorption–desorption isotherms of the synthesized samples are shown in Fig. 5. The isotherms of all samples followed type I and IV according to the IUPAC classification, which indicate the presence of both microporosity and mesoporosity in all prepared samples. Steep uptake of the physisorption isotherms of all samples at very low relative pressure of $P/P_0<0.1$ is related to the pore filling of narrow micropores. Moreover, all PEG-directed SAPO-34 materials displayed a hysteresis loop at relative pressures of $P/P_0 > 0.45$, which elucidate the secondary capillary condensation process taking place in the mesopores. These results confirm the hierarchy of synthesized structures with the aid of PEG linear polymers.

Pore shape affects the mechanisms of condensation and evaporation and consequently shape of hysteresis loop. Shape of hysteresis loop in all samples except the conventional sample is close to H4 type with two nearly horizontal branches over a wide range of $P/P_0$ which has been identified with specific slit-like pores. In another interpretation, type H4 merely verifies the existence of limited amount of mesopores implanted in a matrix of tiny micropores. The type H3 in conventional sample with no limiting adsorption at high $P/P_0$ is indicative of the presence of spaces between plate-like particles giving rise to slit-shaped pores. This shape is usually found on solids with a very wide distribution of pore sizes [41], [42]. Continuous uptake at high relative pressures suggests that this sample is an open surface material that allows the formation of multiple adsorbate layers by increasing the $p/p_0$ ratio.

The corresponding BET surface areas, external surface areas, micropore and mesopore volumes are summarized in Table 3. As it can be seen, using PEG in the synthesis gel reduced the BET surface area of all final samples except for the 6-0.0125 and increased their particle sizes. As it has been mentioned before, this increase in particle sizes are probably assigned to the effect of PEG on the viscosity of the precursor wherein the transfer of Si and Al cations to the SAPO zeolitic nucleus were restricted and resulted in lower number of formed nucleus compared to the conventional sample. Also, the attraction between loaded PEG micelles and SAPO patches leads to the adherence and aggregation of huge number of nanocrystals.

The 6-0.0125 sample possesses the highest BET surface area which is the common property of dominant DNL-6 framework with large α cages. The high BET surface area and micropore volume of this sample are comparable to those reported in literature [31]–[33] indicating the high crystallinity.

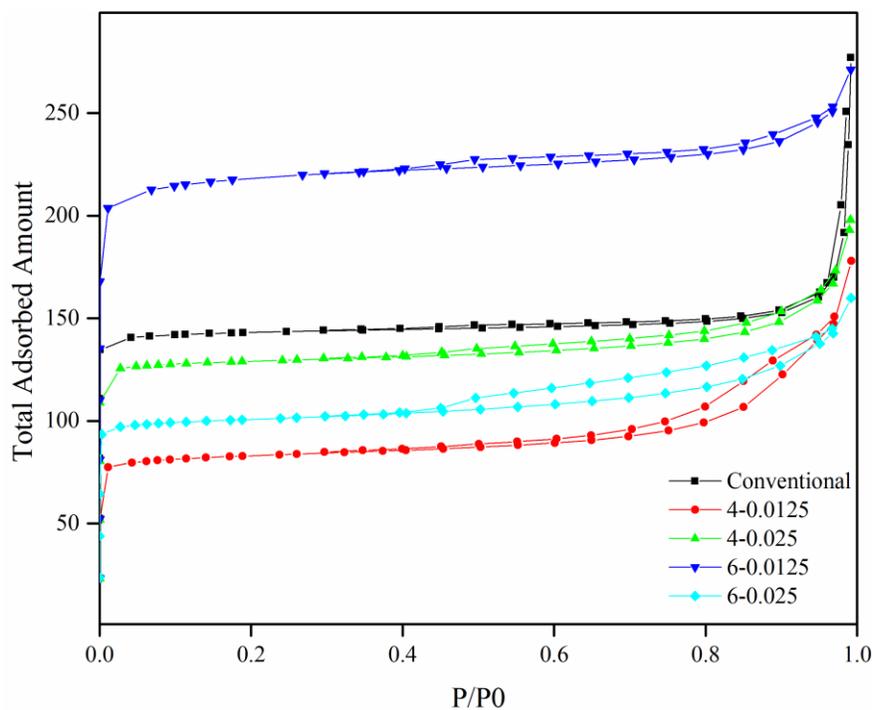

**Fig. 5.** The $N_2$ adsorption–desorption isotherms of the synthesized silicoaluminophosphate zeolites.

**Table 3**

Physicochemical properties of the synthesized silicoaluminophosphate zeolites.

| Sample name | $S_{BET}$ (m2/g)[a] | $S_{EXT}$ (m2/g)[b] | $V_{micro}$ (cm3/g)[b] | $V_{meso}$ (cm3/g)[c] |
|---|---|---|---|---|
| conventional | 589 | 21.7 | 0.233 | 0.158 |
| 4-0.0125 | 340 | 38.0 | 0.096 | 0.174 |
| 4-0.025 | 313 | 46.8 | 0.199 | 0.100 |
| 6-0.0125 | 841 | 30.5 | 0.352 | 0.065 |
| 6-0.025 | 426 | 48.9 | 0.130 | 0.115 |

[a] $S_{BET}$ (total surface area) was calculated using the BET method at 77 K.
[b] $S_{EXT}$ (external surface area) and $V_{micro}$ (micropore volume) were calculated using the t-plot and MP methods.
[c] $V_{meso} = V_t - V_{mic}$ ($V_t$ is calculated using BET method)

Fig. 6 illustrates the BJH and MP pore size distribution of different synthesized SAPO samples in meso and micro scales, respectively. The MP-method is an extension of the t-plot method which is developed by Mikhail and coworkers [43] to obtain the micropore properties from one experimental isotherm. According to the IUPAC classification [39], pores are classified as macropores with pore diameters of greater than 500 Å, mesopores for the pore range 20 to 500 Å and micropores with pore diameters of less than 20 Å. BJH analysis gives good estimation on mesopore size distribution while MP method can effectively predict the micropore size distribution providing a whole scheme of the porosity of each sample. As it can be observed from BJH diagrams, sharp peaks around 3-5 nm exist in all PEG directed samples while random broad humps in meso and macro scales are present in conventional sample. These decentralized peaks are assigned to the intercrystalline gaps created upon crystals intensification in conventional sample. Conventional sample cannot be considered essentially as mesoporous material because these pores generally depend on the size and intensity of packing of individual crystals and are not reproducible. The mesopore volume reported in Table 3 for this sample is the volume of these random spaces between crystals. In fact, PEG inclusion tailored the porosity towards the formation of tuned mesopores.

Moreover, strong single peaks at 0.6 nm are perceived in the MP diagrams of conventional, 4-0.0125 and 4-0.025 samples which are close to the maximum diameter of a sphere that can be included in CHA cages (7.37 Å) (Fig. 7). An intense peak at 0.8 nm in 6-0.0125 sample verifies again the domination of DNL-6 phase with large α-cages of 10.43 Å. The MP pore size distribution of 6-0.025 sample including both CHA and RHO structures showed a strong peak at 0.7 nm; the mean value of nearly pure CHA and RHO frameworks peaks.

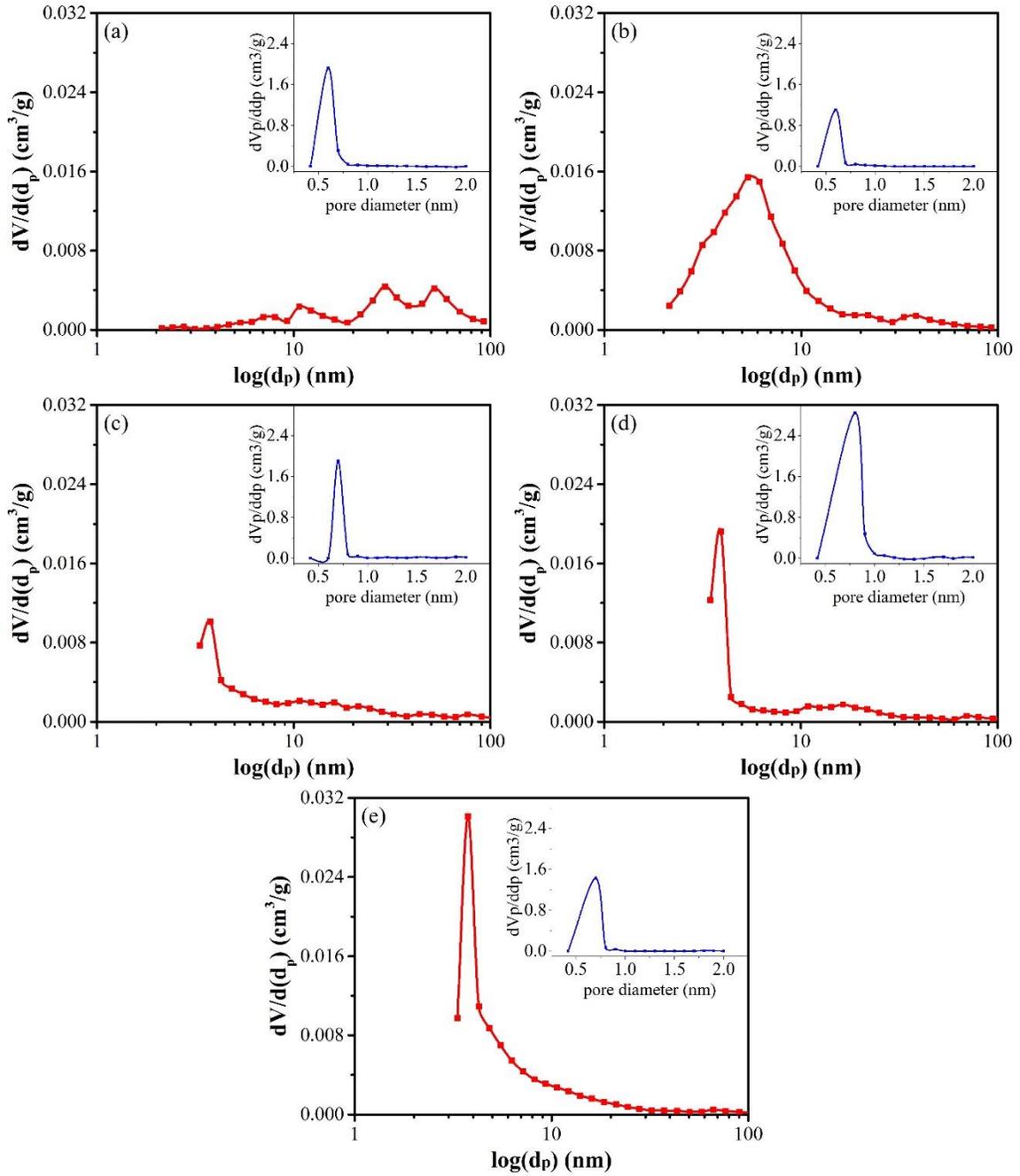

**Fig. 6.** The BJH pore size distributions and the MP plots (inset) corresponding to the desorption branches of a) conventional, b) 4-0.0125, c) 4-0.025, d) 6-0.0125 and e) 6-0.025 samples.

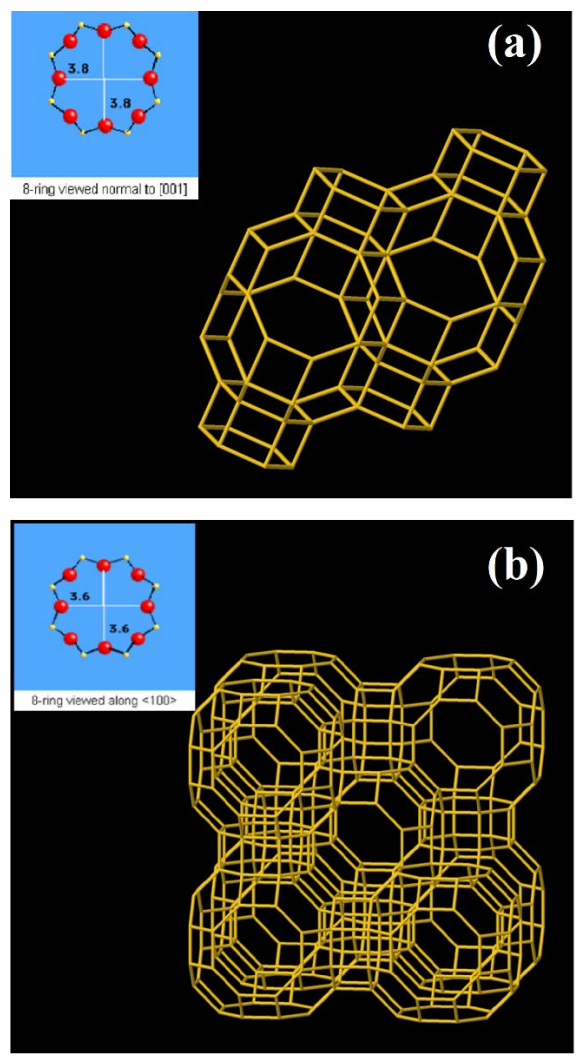

**Fig. 7.** Stereoview of a) CHA structure and b) RHO structure (8-ring window opening of each structure is shown inside the figures).

NH$_3$-TPD was used to investigate the acidic properties of the conventional, 4-0.0125, and 6-0.0125 catalysts, and the results are shown in Fig. 8 and Table 4. The desorption temperature signifies the strength of acid sites, given that the stronger acid sites require a higher desorption temperature. Area under the NH$_3$ desorption curve indicates the amount of ammonia desorbed, which is proportional to the acidity [44]. $NH_3$-TPD profiles of the samples show three distinctive $NH_3$ desorption peaks at around 200, 400, and 600 °C, which implies the presence of acid sites with different level of strengths.

The first peak around 200 °C is presented by the weak acid sites from surface hydroxyl groups bounded to the defect sites such as P–OH, Si–OH, and Al–OH [45], [46]. Although the first peaks areas are different for all samples, the corresponding acid sites could not contribute in conversion of methanol at the reaction temperature of 400 °C. T-OH (T stands for Al, Si, or P atoms placing in the center of the tetrahedron formed from T-O bonds) acid sites, attributed to the first peaks in TPD curves, desorbed ammonia at temperature well below 400 °C. In fact, adsorbed ammonia molecules totally desorbed before 300 °C. Therefore, methanol as a non-basic molecule could not adsorb on these types of acid sites (T-OH) at reaction temperature of 400 °C to be converted to olefins and variable amount of T-OH acid sites represented the by first peaks had no effect on catalytic performance of SAPO-34 samples. The second and third desorption peaks were assigned to the bridging hydroxyl group of –SiOHAl– linkages as strong Bronsted acid sites responsible for strong acidity of SAPO molecular sieves [47]. According to the average desorption temperature of these strong acid sites in TPD profiles, ca. 400 °C and 600 °C, it seems that these kinds of hydroxyl groups would be able to adsorb and convert methanol molecules at the reaction temperature of around 400 °C and free hydroxyl groups (T-OH) related to the first desorption peak in TPD profiles has no role in MTO conversion. So, considering the second and third peaks in ammonia desorption profiles, 4-0.0125 sample possessed the highest acidity. Also, the second ammonia desorption peak in 4-0.0125 sample's profile is slightly higher, indicating the stronger acidity of these sites compared to other materials.

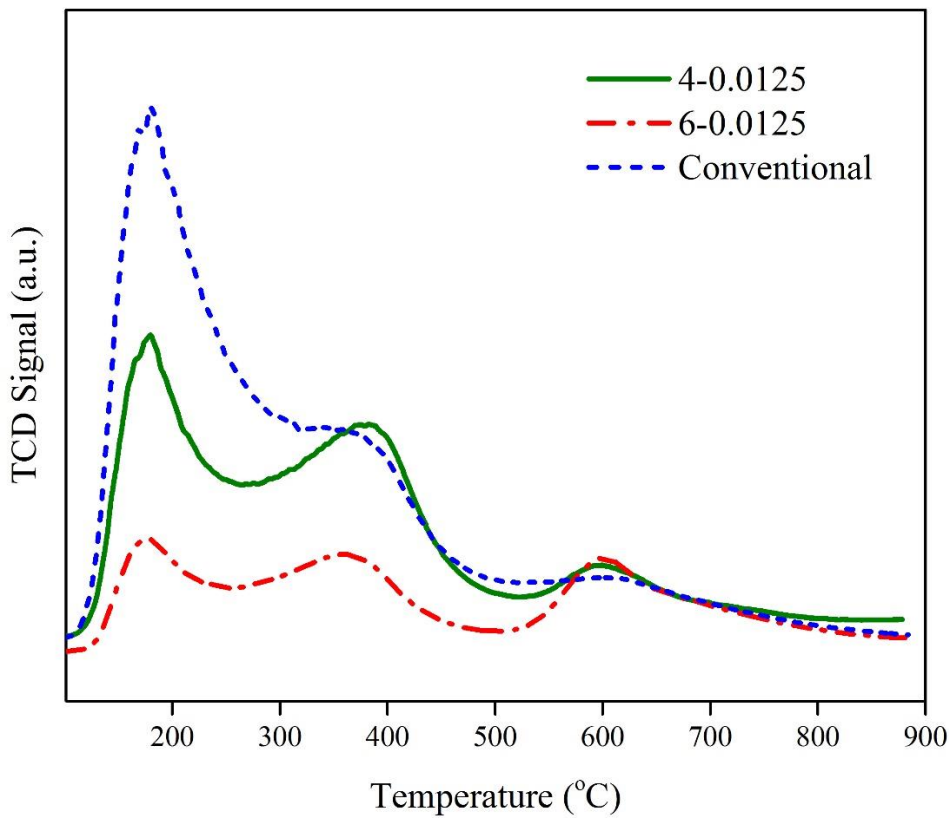

**Fig. 8.** The NH$_3$-TPD profiles of the 4-0.0125, 6-0.0125, and conventional catalysts.

**Table 4**

NH$_3$-TPD analysis of prepared samples.

|  | *Acidity (mmol g-1)* | | | |
| --- | --- | --- | --- | --- |
| sample | weak | moderate | strong | moderate+ strong |
| *Conventional* | 1.007 | 0.463 | 0.208 | 0.671 |
| *4-0.0125* | 0.464 | 0.626 | 0.218 | 0.844 |
| *6-0.0125* | 0.146 | 0.247 | 0.254 | 0.501 |

Table 5 listed the product composition of 4-0.0125 and 6-0.0125 selected samples as representative of dominant CHA and Rho structural phases. In the construction of SAPOs frameworks, the number of Al atoms should be equal to the number of Si plus P atoms due to the absence of Si–O–P linkages. Values of Al/(Si + P) ratio of these two samples approached one, indicating the high purity of crystal products which were consistent with XRD results and rich Si(4Al) environment of Si atoms. Also, the Si incorporation for 6-0.0125 sample is higher than 4-0.0125 sample which is due to the instinctive higher silicon content of DNL-6 crystals.

**Table 5**

Gel and product compositions for the synthesis of 4-0.0125 and 6-0.0125 samples.

| Sample | Gel composition PEG(MW)/Al/P/Si | Product composition Si/Al/P | Al/(Si+P) | Si incorp.$^a$ |
|---|---|---|---|---|
| *4-0.0125* | 0.0125(4000):2.0:1.6:0.5 | 0.13:0.47:0.4 | 0.89 | 1.06 |
| *6-0.0125* | 0.0125(6000):2.0:1.6:0.5 | 0.16:0.45:0.39 | 0.82 | 1.31 |

a Si incorporation= [Si/(Si+Al+P)]$_{product}$/[Si/(Si+Al+P)]$_{gel}$.

MTO reaction tests were applied to determine the catalytic performance as well as the life time of different silicoaluminophosphate samples, during reaction. The results of methanol conversion plots versus time on stream are shown in Fig. 9. The intermediate dimethyl ether, as the condensation product of methanol, is also regarded as unconverted component. In MTO reaction, high acidity gives rise to a rapid deactivation and low selectivity to light olefins despite high catalytic activity because of the acceleration of cracking side reaction. All catalysts revealed full conversion of methanol at the beginning of the process except 6-0.0125 sample which is mostly composed of DNL-6; a low acidic SAPO counterpart according to the table 5. However, except for the 4-0.0125 sample, methanol conversion decreased after 150-200 min drastically until no conversion was detected anymore. The 4-0.0125 showed a full conversion of methanol and dimethyl ether for almost 320 min time on stream despite the higher effective acidity compared to the conventional sample. Additionally, the deactivation rate was slower in 4-0.0125 sample rather than the other samples. The improved reactivity, especially catalytic durability of 4-0.0125 catalyst was induced by the tailored mesoporosity and effective acidity of this material. The introduced wide mesopores acted as auxiliary diffusion pathways and they led to a better accessibility of the acidic sites and higher catalyst efficiency. These pores are a solution to the bottleneck of very low diffusion of products in narrow micropores which can be stuck at the body or mouth of the pores leading to the blockage and deactivation of catalyst.

Hydrocarbon pool (HCP) mechanism has been proposed in recent years as leading mechanism in MTO reaction wherein HCP species (polyaromatic compounds) accumulate within zeolite cages [48]–[50]. This mechanism is an indirect catalytic cycle possessing lower energy barriers compared to the direct formation of C−C bonds from C1 moieties [51]–[53]. In this mechanism, HCP species function as scaffolds (reaction centers) for C−C bond breaking and formations to produce desired olefins. Hexa/pentamethylbenzene (H/PMB) are the most important and reactive reaction centers among polymethylbenzene intermediates formed in SAPO-34 cavities of about 7 Å through ship in a bottle route [54], [55]. The incoming methanol reacts with these intermediates through side-chain methylation reaction and olefins are released through rearrangement and dealkylation reaction. Larger cages of DNL-6 structure (∼1 nm) in 6-0.0125 and 0.025 samples can normally take up and accommodate several cyclic intermediate compounds. Because of narrower openings of this structure compared to SAPO-34 material, methanol accessibility to these aromatic compounds is postponed and polymethylbenezene age into methylnaphtalenes and then phenanthrene which are less reactive high carbonaceous compounds. These substances are believed to restrict the contact of methanol with reaction centers and ultimately cause the loss of activity as the polycyclic coke species [56]. So, conversion is dropped more quickly in both DNL-6/SAPO-34 composite catalysts.

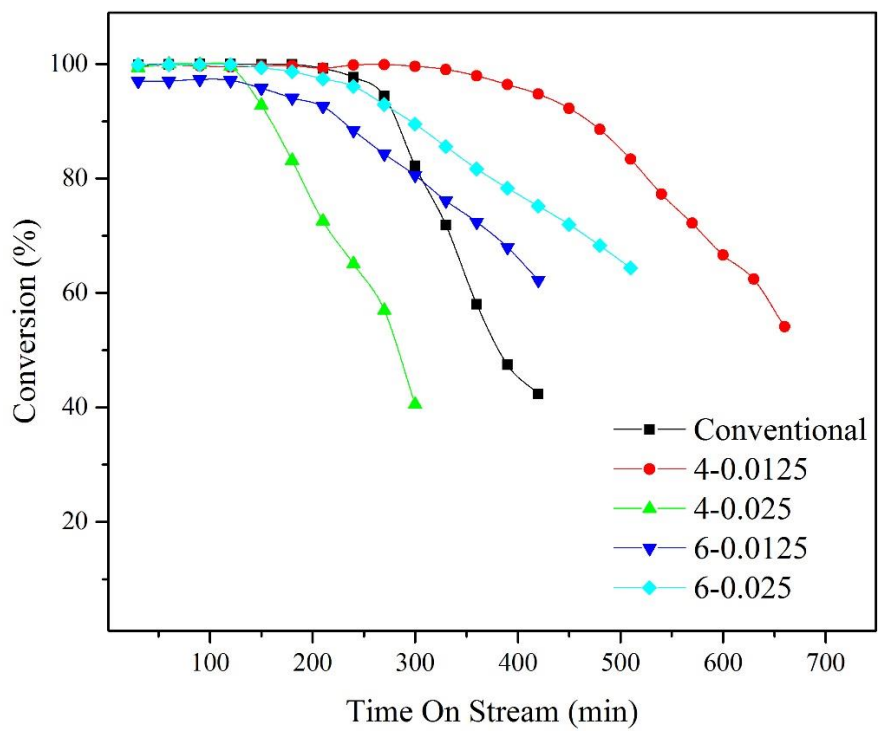

**Fig. 9.** Methanol and dimethyl ether conversions of the synthesized silicoaluminophosphate zeolites. Experimental conditions: WHSV = 3 h$^{-1}$, T = 673 K, catalyst weight = 500 mg.

The MTO product selectivities of the samples are shown in Fig. 10 a–f. All catalysts showed a similar product distribution, with ethylene and propylene as the main products. The selectivity is a result of the size and shape sieving effect due to the narrow pore openings of the SAPO-34 and DNL-6. Only very low amounts of C5-molecules and no aromatic products were observed. For all samples, the ethylene and propylene selectivities initially increased with time on stream, which can be subscribed to persistency of catalyst towards coke formation as it is reported in the literature [5], [57]. The rough similarity of the product spectra confirms that the reaction centers are the same. After the deactivation of the catalysts, dimethyl ether was formed by methanol condensation in large amounts. The absence of any aromatic or bulky molecules indicates that for all samples, the reaction centers are inside the molecular sieve cavities and the introduction of mesopores does not change the product formation significantly. This issue has been proven by several authors in the literature previously [58]–[60].

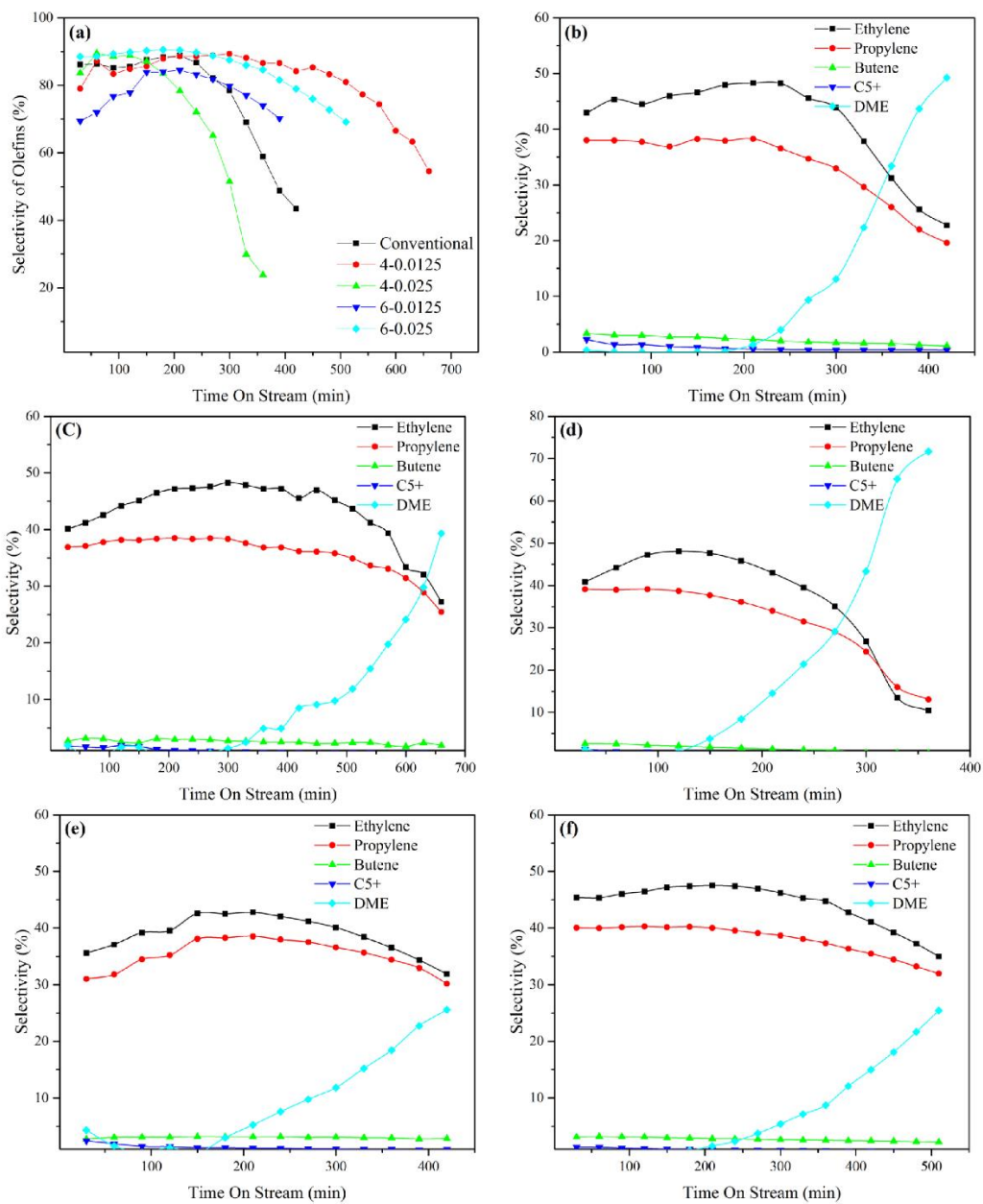

**Fig. 10.** a) Total light olefins selectivity and Product selectivity of the synthesized silicoaluminophosphate zeolites: (b) conventional; (c) 4-0.0125; (d) 4-0.025; (e) 6-0.0125; and (f) 6-0.025. Experimental conditions: WHSV = 3 h$^{-1}$, T = 673 K, catalyst weight = 500 mg.

The activity of regenerated 4-0.0125 catalyst as the optimum catalyst is reported in Fig. 11. After deactivation occurred, synthetic air was passed through the catalyst bed at a temperature of 823 K for 2 h. Subsequently, the temperature was set to 673 K again and the catalytic reaction was replicated under the same conditions. After six regeneration cycles, the catalyst represented at least 96% of the initial conversion capacity. This confirms that the SAPO-34 retains its regeneration stability even after functionalization with additional hierarchical pore architecture.

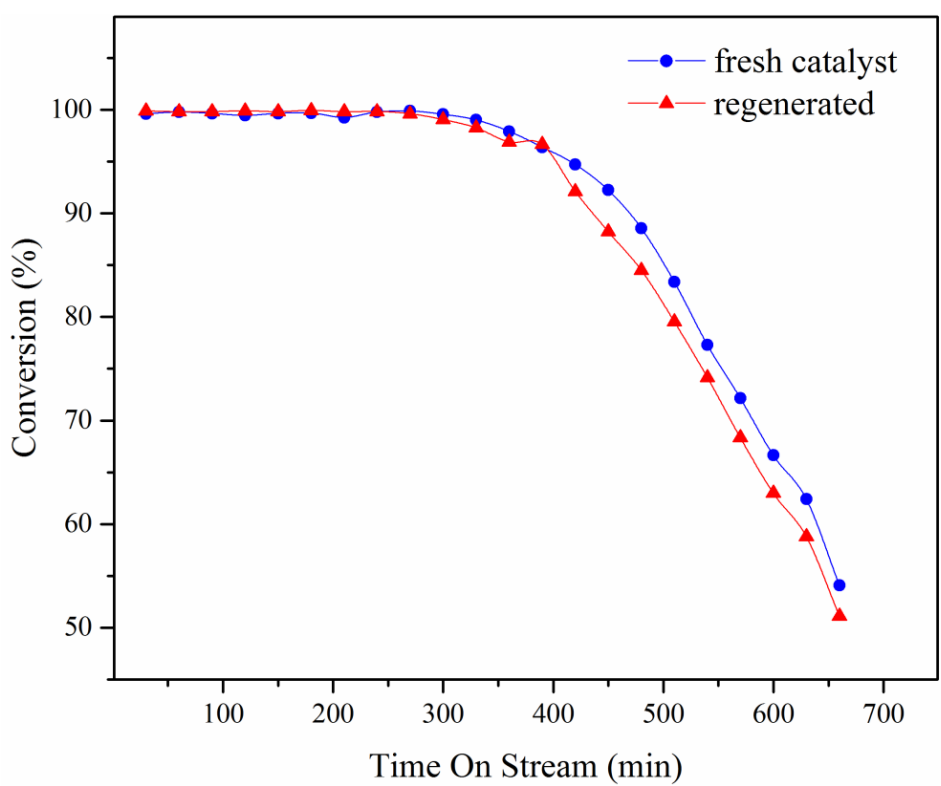

**Fig. 11.** Catalytic lifetime of the fresh and six-times regenerated 4-0.0125 sample. Experimental conditions: WHSV = 3 h$^{-1}$, T = 673 K, catalyst weight = 500 mg.

**Conclusion**

In this study, silicoaluminophosphate-34 zeolites with microporous–mesoporous hierarchical tailored structures were successfully synthesized using combination of amine agents as template and PEG as the mesogenerating agent. Large well-crystallized but less symmetric particles were obtained upon the introduction of PEG organic additive to the synthesis media. This was observed because of the effect of PEG on the viscosity and surface tension of the precursor wherein the transfer of cationic structural elements to the SAPO zeolitic nucleus were restricted and the attraction between loaded PEG micelles and SAPO patches was led to the adherence and aggregation of huge number of nanocrystals. The effect of PEG molecular weight and PEG/Al molar ratio on the compositional and textural properties of the synthesized materials were investigated in order to optimize their MTO performances. PEG/Al Molar ratio of the precursor significantly affected the crystallinity, porosity and acidity of the synthesized products. Using PEG with MW of 4000 has resulted in pure SAPO-34 with CHA structure while PEG with MW= 6000 has led to the formation of DNL-6 secondary phase with RHO framework next to SAPO-34. PEG with MW=4000 templated SAPO-34 with PEG/Al molar ratio of 0.0125 showed full conversion, ∼90% selectivity towards light olefins and superlative catalytic stability in MTO process because of the highest degree of mesoporosity and tailored acidity pattern. Similar light olefins selectivity spectra of all catalysts confirm this fact that the established mesopores did not alter the products formation pattern and were only served as diffusion pathways for better accessibility of active sites located within intersections and cavities.